\documentclass[aps,groupedaddress,showpacs,letterpaper,twocolumn,reprint,nofootinbib,longbibliography]{revtex4-1}
\usepackage{graphicx} 
\usepackage{amsmath} 
\usepackage{amssymb}   
\usepackage{color}  

\newcommand{\ket}[1]{| #1 \rangle}
\newcommand{\bra}[1]{\langle #1 |}

\begin{document}

\title{Rydberg blockade, F\"orster resonances, and quantum state measurements with different atomic species}
\author{I. I. Beterov}
\affiliation{Rzhanov  Institute of Semiconductor Physics SB RAS, 630090 Novosibirsk, Russia}
\affiliation{Novosibirsk State University, Quantum Center, 630090 Novosibirsk, Russia}
\author{M. Saffman }
\affiliation{
Department of Physics,
University of Wisconsin-Madison, 1150 University Avenue,  Madison, Wisconsin 53706
}
 \date{\today}

\begin{abstract}
We calculate interspecies Rydberg-Rydberg interaction strengths for the heavy alkalis Rb and Cs. The presence of strong F\"orster resonances makes interspecies coupling a promising approach for long range entanglement generation. 
We also provide an overview of the strongest  F\"orster resonances for Rb-Rb and Cs-Cs using different principal quantum numbers for the two atoms. We show how interspecies coupling can be used for high fidelity quantum non demolition state measurements with low crosstalk in qubit arrays. 
\end{abstract}

\maketitle

\section{Introduction}

Optically trapped neutral atoms are being actively developed for quantum simulation and quantum computing  
applications\cite{Mueller2012,Saffman2010}
and there has been substantial recent progress in improving the fidelity of one- and two-qubit gate operations\cite{Xia2015,YWang2015,Anderson2015,Jau2015,Maller2015a}.
Several different approaches are possible for encoding qubits in neutral atoms. For example collective encoding provides a method for establishing a multi-qubit   register in the collective states of a single atomic ensemble\cite{Brion2007d}. One of the challenges in implementation of  collective encoding is measuring the  state of a single qubit without disturbing the rest of the register. This can in principle be done by state selective excitation to a Rydberg level followed by ionization. This has the drawback of suffering from less than unity 
quantum efficiency  of practical ion detectors, plus the problem of atom loss. After each measurement of a bit value of $|1\rangle$ an atom is lost and has to be replaced from the collective reservoir state. The number of measurements which can be made before the reservoir is depleted is thus limited by the number of atoms in the ensemble.  
An alternative  is to perform a Rydberg gate between the register to be measured and an auxiliary register (or single qubit) in a neighboring trap. The state of the auxiliary bit can then be measured without atom loss.  This has the drawback of requiring a longer range 
gate to be performed. For qubits encoded in a single atom, optical trap arrays can be used to define a multi-qubit register\cite{Piotrowicz2013,Xia2015,Nogrette2014,YWang2015}. Also in this case measurement of the state of a single qubit without disturbance of proximal qubit locations is challenging due to the isotropic distribution of light scattered during a measurement.

\begin{figure}[!t]
\centering
\includegraphics[width=6.cm]{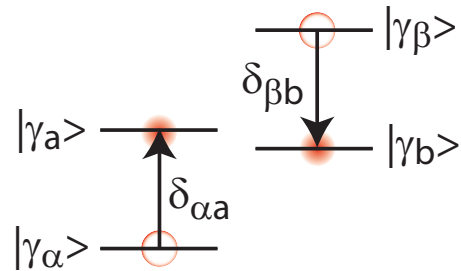}
\caption{(color online) The resonance condition for interspecies dipole-dipole coupling between initial states $a,b$ and target states $\alpha,\beta$ is $\delta_{\alpha a}=\delta_{\beta b}.$ }
\label{fig.1}
\end{figure}

State measurements may also be based on cross entanglement of two different atomic elements located in the same trap, or nearby traps. By creating entanglement between qubits encoded in different types of atoms the quantum state of a qubit encoded in atom $a$ can be measured via light scattering from the qubit encoded in atom $b$. This is analogous to the mixed species quantum logic spectroscopy previously demonstrated with trapped ions\cite{Schmidt2005}. For example   Rb atoms have D1 and D2 resonance lines at 795 and 780 nm while Cs atoms have D1 and D2 lines at 894 and 852 nm. The large separation implies that measurements, as well as optical pumping and state preparation, can be performed independently on nearby atoms. 
Atoms of different species $a$ and $b$ can have a strong dipole-dipole interaction due to a F\"orster type mechanism 
when the energy defect $\hbar\delta=\hbar(\delta_{\alpha a}+\delta_{\beta b})= (U_\alpha-U_a)+(U_\beta-U_b)$ is small as shown in Fig. \ref{fig.1}. Here $a,b$ denote initial quantum states and $\alpha,\beta$ the dipole coupled states.  In this paper we provide a detailed analysis of interspecies F\"orster resonances for Rb and Cs atoms and analyze the application of the interspecies coupling to quantum non demolition (QND) state measurements.

The structure of the paper is as follows. In Sec. \ref{sec.theory} we provide general formulae for calculating interspecies dipole-dipole interactions. The formalism generalizes the results of \cite{Walker2008} to the situation where the laser excited atoms are not in the same quantum state. In Sec. \ref{sec.RbCs} we present a list of useful 
F\"orster resonances for Rb-Cs coupling. In Sec. \ref{sec.RbRb}  we list the strongest resonances for coupling Rb-Rb and Cs-Cs using different principal quantum numbers $n$ for each atom. In Sec. \ref{sec.angular} the angular variation of the interaction is calculated for isotropic, and strongly anisotropic cases, and in Sec. \ref{sec.measurement} we discuss the problem of qubit measurement and show how the interspecies coupling can be used for fast measurements with very low crosstalk. Section \ref{sec.summary} summarizes our results.

\begin{table*}[!t]
\caption{Interaction channels $k$, eigenvalues $D_{k\ell}$, and eigenvectors $\ket{u_{k\ell}}$   for atoms in  $ns_{1/2}$ states.   Eigenvalues $D_{k\ell}$ are for atoms in identical initial states $\gamma_a\ne\gamma_b$ with allowed  couplings 
$(ab)\leftrightarrow(\alpha\beta)$. Eigenvalues $D_{k\ell}'$ 
are for atoms in initial states $\gamma_a=\gamma_b$ with allowed  couplings 
$(ab)\leftrightarrow(\alpha\beta)$ and $(ab)\leftrightarrow(\beta\alpha)$  .    The eigenvectors specified in terms of states $\ket{m_a,m_b}$ are the singlet and triplet states 
$\ket{u_s}=\frac{1}{\sqrt2}(\ket{1/2,-1/2}-\ket{-1/2,1/2})$, $\ket{u_{t0}}=\frac{1}{\sqrt2}(\ket{1/2,-1/2}+\ket{-1/2,1/2})$, $\ket{u_{t\pm}}=\ket{\pm1/2,\pm1/2}$. States $s,t0,t\pm$ are labeled as $\ell=1,2,3,4$ respectively. }
\vspace{-.4cm}
\begin{center}
\begin{ruledtabular}
\begin{tabular}{c|c|c|c|c|c|c}
channel $k$& $j_{\alpha }$ & $j_{\beta }$ &$m=m_a+m_b$& $\ket{u_{k\ell}}$&$D_{k\ell}$&$D_{k\ell}'$\\ 
\hline
\hline
 1& $1/2$ & $1/2$ &0&$\ket{u_s}$& $0$&$0$\\
   &  &  &0& $\ket{u_{t0}}$&$16/9$&$32/9$\\
  &  &  & $\pm1$&$\ket{u_{t\pm}}$&$4/9$&$8/9$\\
\hline
\hline
  2& $1/2$ & $3/2$ &0&$\ket{u_s}$& $2$&$4$\\
 &  &  &0&$\ket{u_{t0}}$& $2/9$&$4/9$\\
   &  &  & $\pm1$&$\ket{u_{t\pm}}$&  $14/9$&$28/9$\\
\hline
\hline
3& $3/2$ & $1/2$ &0&$\ket{u_s}$& $2$&$4$\\
&  &  &0& $\ket{u_{t0}}$&$2/9$&$4/9$\\
&  &  & $\pm1$&$\ket{u_{t\pm}}$&$14/9$&$28/9$\\
\hline
\hline
4& $3/2$ & $3/2$ &0&$\ket{u_s}$& $2$&$4$\\
&  &  &0& $\ket{u_{t0}}$&$34/9$&$68/9$\\
&  &  & $\pm1$&$\ket{u_{t\pm}}$&$22/9$&$44/9$\\

\end{tabular}
\end{ruledtabular}
\end{center}
\label{tab.Dk}
\end{table*}

\section{Dipole-Dipole Rydberg interaction between distinguishable atoms}
\label{sec.theory}

In this section we provide explicit expressions for calculating the interspecies dipole-dipole interaction between atoms $a$ and $b$ leading to Rydberg blockade. Our notation mostly follows the theory of \cite{Walker2008} with some  modifications, and slightly generalized to allow for the  initial Rydberg pair states  to be distinguishable.
We characterize the strength of the interaction for a particular angular momentum channel by the $C_3$ and van der Waals coefficients. 
The label $\gamma_a=(z_a,n_a,l_a,j_a)$  denotes the quantum numbers of a single Rydberg level $a$.  The  coupling 
$(ab)\leftrightarrow (\alpha\beta)$ specifies an interaction channel $k$ coupling a pair of atoms in fine structure levels $a,b$ to a pair of atoms in fine structure levels $\alpha ,\beta $. Here $z$ specifies the atomic species, $n$ is the principal quantum number, $l$ is the orbital angular momentum, and $j$ is the total electronic angular momentum of a fine structure state. We assume single electron $s=1/2$ atoms throughout.

   We define the  $C_{3,k}$ coefficient of  channel $k$ as
\begin{equation}
C_{3,k}(a,b,\alpha,\beta)=q^2\frac{\langle\gamma_{\alpha }||r_a||\gamma_a \rangle
\langle\gamma_{\beta }||r_b||\gamma_b\rangle}{\sqrt{(2j_{\alpha }+1)(2j_{\beta }+1)}},
\label{eq.C3k}
\end{equation}
with $q^2=e^2/4\pi\epsilon_0$, $e$ is the electronic charge, $\epsilon_0$ is the permittivity of free space, and  $\langle\gamma_{\alpha }||r_a||\gamma_a\rangle$ is a reduced matrix element in the fine structure basis. 
This differs from the notation of  \cite{Walker2008} where the $C_3$ coefficient was defined in terms of radial matrix elements in the $n,l$ basis. Note that $C_{3,k}$ depends on a total of 14 parameters:
$z_a, z_b, n_a, l_a, j_a, n_b, l_b, j_b, n_{\alpha }, l_{\alpha },j_{\alpha },
n_{\beta }, l_{\beta }, j_{\beta }$.

The energy defect for channel $k$ is 
$\hbar\delta_k=\hbar(\delta_{\alpha a}+\delta_{\beta b})= [U(\gamma_{\alpha })-U(\gamma_a)]+[U(\gamma_{\beta })-U(\gamma_{b})]$.
In the approximation that a single channel dominates the interaction  the energy shift of a F\"orster eigenstate  $\ket{u_{k\ell}} $ depends on the interatomic separation $R$ as 
\begin{equation}
U_{k\ell}(m_a,m_b) = \frac{\hbar\delta_k}{2}\left[1- \left(1+\frac{4D_{k\ell}(m_a,m_b)C_{3,k}^2}{\hbar^2\delta_k^2R^6}\right)^{1/2}\right].
\label{eq.Uk}
\end{equation}
The angular factor $D_{k\ell}$ is always positive so for  $\delta_k>0 (<0)$ the interaction is attractive(repulsive). The long range van der Waals interaction for eigenstate $\ell$ in channel $k$ is 
\begin{equation}
U_{k\ell, \rm vdW}=-\frac{D_{k\ell}C_{3,k}^2}{\hbar\delta_k}\frac{1}{R^6}.\nonumber
\end{equation}
We define a crossover distance $R_c$ marking the boundary between a $1/R^3$ resonant interaction and a $1/R^6$ van der Waals interaction 
by
$$
R_c=\left(\frac{D_{k\ell}C_{3k}^2}{\hbar^2\delta_k^2}\right)^{1/6} .
$$
The angular factor $D_k(m_a,m_b)$ depends on the quantum numbers of the interacting states and is  calculated with the method described  in Appendix A. 

When we consider the interaction of atoms of different types,  either two different atomic elements, or two different isotopes of one element, we have $\gamma_a\ne \gamma_b$ and only include the coupling $(ab)\leftrightarrow(\alpha\beta)$. Also for atoms of the same type but with $\gamma_a\ne \gamma_b$ there will usually  only be  a single coupling $(ab)\leftrightarrow(\alpha\beta)$ which is dominant. The $D_{k\ell}$ values for channel $k$ and eigenvector $\ell$ for   $ns_{1/2}$ states  are given in Table \ref{tab.Dk}. Interaction of atoms of the same type which are prepared in the same levels, $\gamma_a=\gamma_b$, will have two sets of couplings of the same strength: $(ab)\leftrightarrow(\alpha\beta)$ and $(ab)\leftrightarrow(\beta\alpha)$. This gives the twice larger $D_{k\ell}'$ values given in Table \ref{tab.Dk}, which are in agreement with the values given in Table I of \cite{Walker2008}\footnote{To compare the values in Table I of \cite{Walker2008} with those given here it is necessary to account for the different definitions of $C_3$.}. 

Starting with a specific molecular Rydberg state $\ket{\psi}=\sum_{ij}c_{ij}\ket{m_{ai},m_{bj}}$ the interaction energy due to channel $k$ is found by decomposing into the F\"orster eigenstates $\ket{u_{k\ell}}$. Writing $\ket{\psi}=\sum_\ell c_{k\ell} \ket{u_{k\ell}}$ with $c_{k\ell}=\bra{u_{k\ell}}\psi\rangle$ we have
\begin{equation}
U_{\ket{\psi},k}=\sum_\ell |c_{k\ell}|^2 U_{k\ell}.\nonumber
\end{equation}

When there are multiple interaction channels $\{k\}$, corresponding to additional values of $\gamma_\alpha, \gamma_\beta,$ the situation is more complicated and in general has to be treated by  numerical solution of the eigensystem of the matrix in Eq. (\ref{eq.M}), extended to include multiple channels. 
When $R\gg R_c$ so the interaction energy is small compared to the F\"orster energy defect there is negligible amplitude of the target states $\alpha,\beta$ and in a first approximation we may assume  the energy shifts are additive. In this van der Waals limit  the interaction energy is 
\begin{equation}
U_{\ket{\psi},\rm vdW}=\sum_{k,\ell} |c_{k\ell}|^2  U_{k\ell,\rm vdW}.
\label{eq.UpsivdW}
\end{equation} 
At small $R$ where the interaction is resonant and there is substantial state mixing we must  account for coupling between channels, which is most conveniently done numerically. The interchannel coupling may lead to nonadditive behavior, as has been discussed previously\cite{Pohl2009,Cano2012}

\section{Rb-Cs F\"orster resonances}
\label{sec.RbCs}

\begin{figure}[!tb]
\centering
\includegraphics[width=.99\columnwidth]{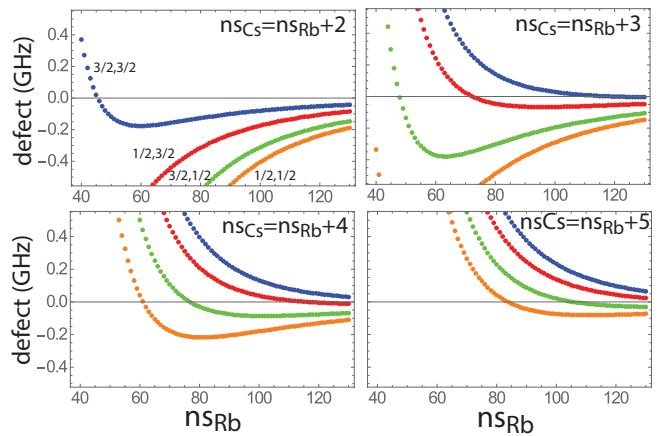}
\caption{(color online) Energy defects for interspecies coupling with $ns_{Cs}> ns_{Rb}$, $np_{\rm Rb}=ns_{\rm Rb}$,
 $np_{\rm Cs}=ns_{\rm Cs}$-1. The different curves show the  $(j_{p \rm Rb},j_{p \rm Cs})$ channels in the sequence $(3/2,3/2), (1/2,3/2), (3/2,1/2), (1/2,1/2)$ from top to bottom.  }
\label{fig.Rb_Cs_defects1}
\end{figure}

\begin{figure}[!tb]
\centering
\includegraphics[width=.99\columnwidth]{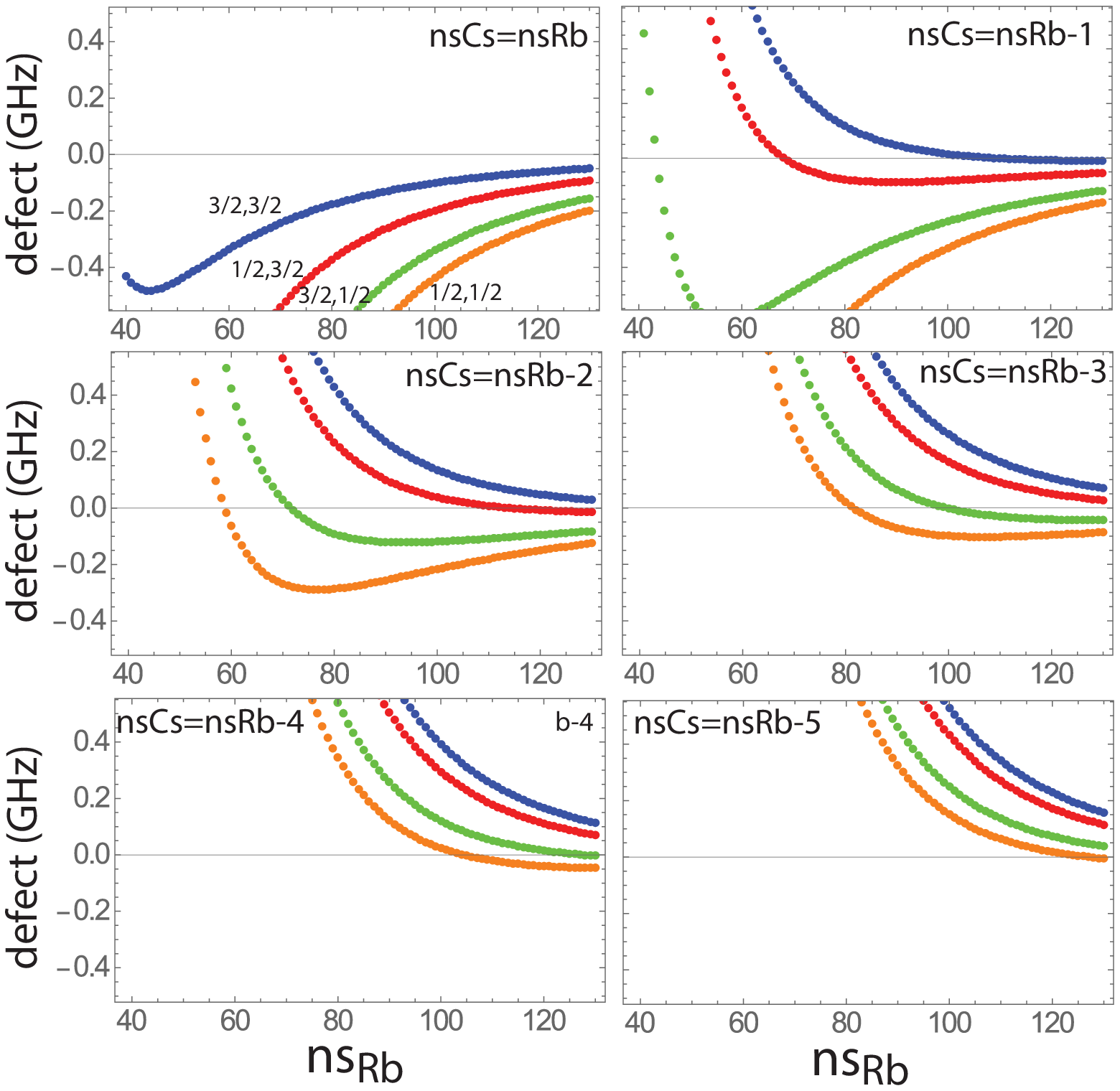}
\caption{(color online) Energy defects for interspecies coupling with $ns_{Cs}\le ns_{Rb}$, $np_{\rm Rb}=ns_{\rm Rb}-1$,
 $np_{\rm Cs}=ns_{\rm Cs}$. The different curves show the  $(j_{p, \rm Rb},j_{p, \rm Cs})$ channels in the sequence $(3/2,3/2), (1/2,3/2), (3/2,1/2), (1/2,1/2)$ from top to bottom.  }
\label{fig.Rb_Cs_defects2}
\end{figure}

F\"orster resonances  for Rb-Cs coupling occur for  a range of angular momentum channels. 
The simplest  case is excitation of $ns_{1/2}$ levels which are dipole coupled to $np_{1/2}, np_{3/2}$.  Figures \ref{fig.Rb_Cs_defects1}, \ref{fig.Rb_Cs_defects2} show the energy defects  for all four fine structure channels. There are a large number of resonances with the Rb principal quantum number either larger than or smaller than that of Cs. 
Table 
\ref{tab.forster} lists the strongest resonances for $40<n_{\rm Rb}<90.$
 Radial matrix elements were calculated using the WKB approximation of \cite{Kaulakys1995} with quantum defect values taken from Refs. \cite{Li2003Rydberg,Mack2011} for $^{87}$Rb and \cite{Lorenzen1984,Weber1987} for Cs. 

The strongest resonance in the table (last row) provides an interaction strength of 2 MHz at $R=20~\mu\rm m$. Even stronger resonances are available at higher $n$. For example the resonance at 
$n_{\rm Rb}=121, n_{\rm Cs}=124$
gives MHz scale interaction strengths at $R= 45~\mu\rm m$. Note that the energy defect at a resonance can be either positive or negative so the interaction can be either attractive or repulsive. This behavior is distinct from the intraspecies coupling for Rb-Rb or Cs-Cs excited to the same $ns$ states for which the interaction is always repulsive
(see Fig. \ref{fig.Rb_Cs_defects}).

 \begin{table*}[!t]
\caption{F\"orster resonances for Rb-Cs  $ns_{1/2}$ states. Only resonances with defects less than 0.0005 times the level spacing and $|C_{3,k}|>1.0 ~{\rm GHz}\mu\rm m^3$  for $40<n_{\rm Rb}<90$ are listed. 
 The van der Waals interaction strength and $R_c$ distance are listed for the strongest eigenvector for each channel, which are $\ket{u_{t0}}$ for channels 1,4 and  $\ket{u_{s}}$ for channels 2,3.  \label{tab.forster}}
    \begin{ruledtabular}
     \begin{tabular}{cc c c c| c c c c}
channel&Rb$\ket{\gamma_a}$ & Rb$\ket{\gamma_\alpha}$& Cs$\ket{\gamma_b}$ & Cs$\ket{\gamma_\beta}$& $\delta/2\pi$ (MHz) & $C_{3,k}~({\rm GHz} ~\mu\rm m^3)$ &  $U_{k,\rm vdW}~({\rm GHz} ~\mu\rm m^6)$ &$R_c~(\mu\rm m)$\\
\hline

4&$45s_{1/2}$ & $45p_{3/2}$ & $47s_{1/2}$  & $46p_{3/2}$ &28.0&-1.29&-223&4.47\\

4&$46s_{1/2}$ & $46p_{3/2}$ & $48s_{1/2}$  & $47p_{3/2}$ &-11.5&-1.41&656&6.21\\


3&$48s_{1/2}$ & $48p_{3/2}$ & $51s_{1/2}$  & $50p_{1/2}$ &-5.71&1.69&994&7.47\\

1&$59s_{1/2}$ & $58p_{1/2}$ & $57s_{1/2}$  & $57p_{1/2}$ &-16.6&-3.54&1350&6.58\\

1&$61s_{1/2}$ & $61p_{1/2}$ & $65s_{1/2}$  & $64p_{1/2}$ &2.65&-4.80&-15500&13.4\\

2&$68s_{1/2}$ & $67p_{1/2}$ & $67s_{1/2}$  & $67p_{3/2}$ &5.25&6.5&-16100&12.1\\
2&$69s_{1/2}$ & $68p_{1/2}$ & $68s_{1/2}$  & $68p_{3/2}$ &-7.40&6.92&12900&11.0\\

3&$71s_{1/2}$ & $70p_{3/2}$ & $69s_{1/2}$  & $69p_{1/2}$ &9.35&8.01&-13700&10.7\\
3&$72s_{1/2}$ & $71p_{3/2}$ & $70s_{1/2}$  & $70p_{1/2}$ &-7.99&8.51&18100&11.5\\

2&$72s_{1/2}$ & $72p_{1/2}$ & $75s_{1/2}$  & $74p_{3/2}$ &4.61&9.65&-40400&14.4\\
2&$73s_{1/2}$ & $73p_{1/2}$ & $76s_{1/2}$  & $75p_{3/2}$ &-4.31&10.2&48400&15.0\\

3&$77s_{1/2}$ & $77p_{3/2}$ & $81s_{1/2}$  & $80p_{1/2}$ &-2.19&12.3&138000&20.0\\

1&$81s_{1/2}$ & $80p_{1/2}$ & $78s_{1/2}$  & $78p_{1/2}$ &6.31&-13.4&-50800&14.2\\
1&$82s_{1/2}$ & $81p_{1/2}$ & $79s_{1/2}$  & $79p_{1/2}$ &-6.41&-14.2&55600&14.3\\

1&$84s_{1/2}$ & $84p_{1/2}$ & $89s_{1/2}$  & $88p_{1/2}$ &-2.43&-18.2&243000&21.5\\

     \end{tabular}
     \end{ruledtabular}
     \end{table*}

\section{F\"orster resonances of Rb or Cs atoms}
\label{sec.RbRb}

\begin{figure}[!htb]
\centering
\includegraphics[width=.99\columnwidth]{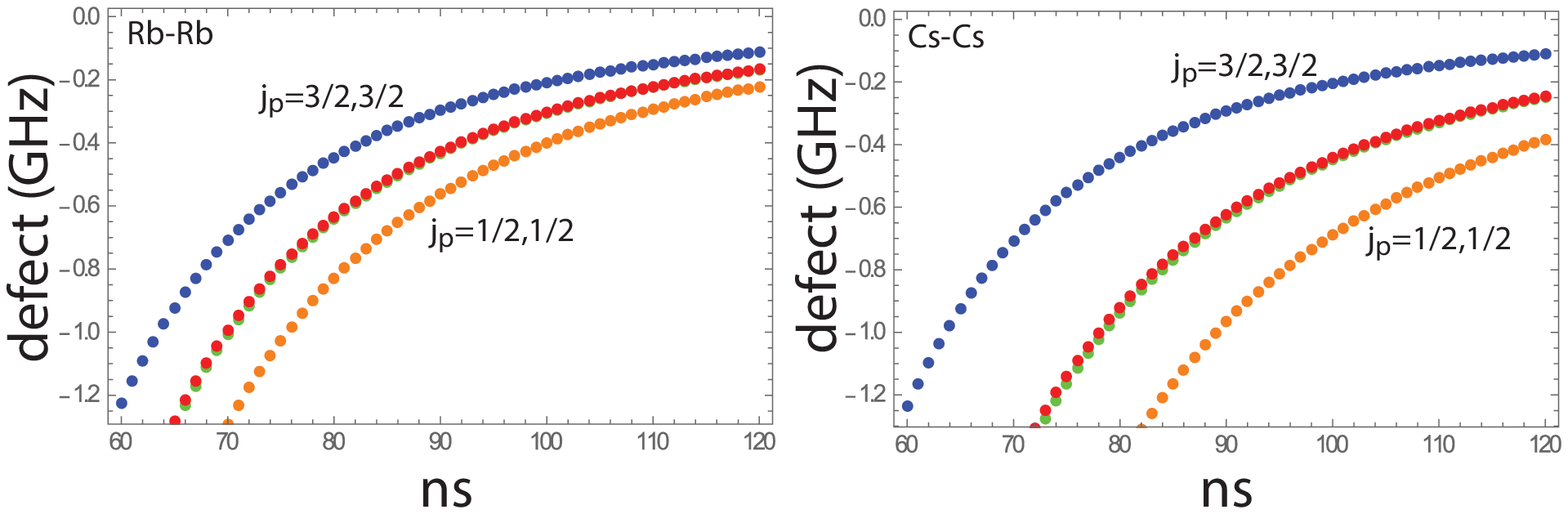}
\caption{Energy defects for intraspecies coupling for the channel $nsns\rightarrow nj_p(n-1)j'_{p}$ for Rb-Rb (left) and Cs-Cs(right). The middle two series in each plot, which are almost identical,  are for $j_p,j'_p=1/2,3/2$ and $3/2,1/2$.    }
\label{fig.Rb_Cs_defects}
\end{figure}

\begin{figure}[!tb]
\centering
\includegraphics[width=.99\columnwidth]{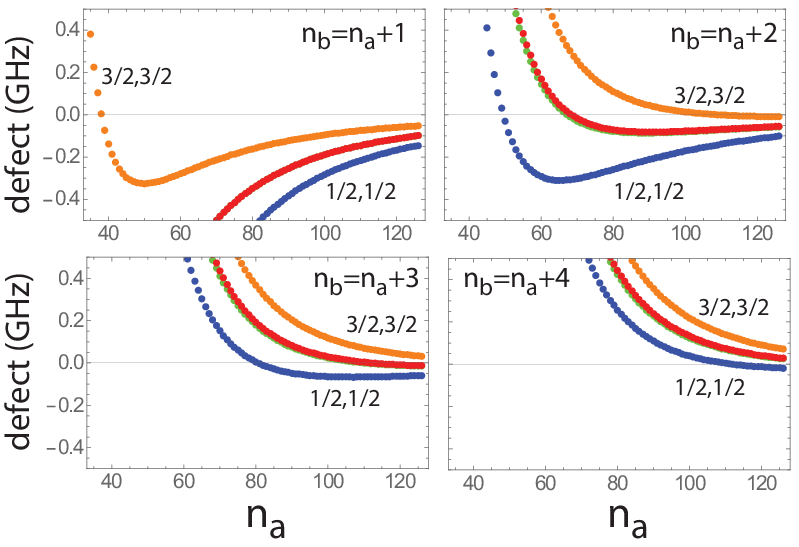}
\caption{(color online) Energy defects for Rb-Rb coupling.  The different curves show the  $(j_{pa},j_{pb})$ channels.  }
\label{fig.Rb_Rb_defects}
\end{figure}

\begin{figure}[!tb]
\centering
\includegraphics[width=.99\columnwidth]{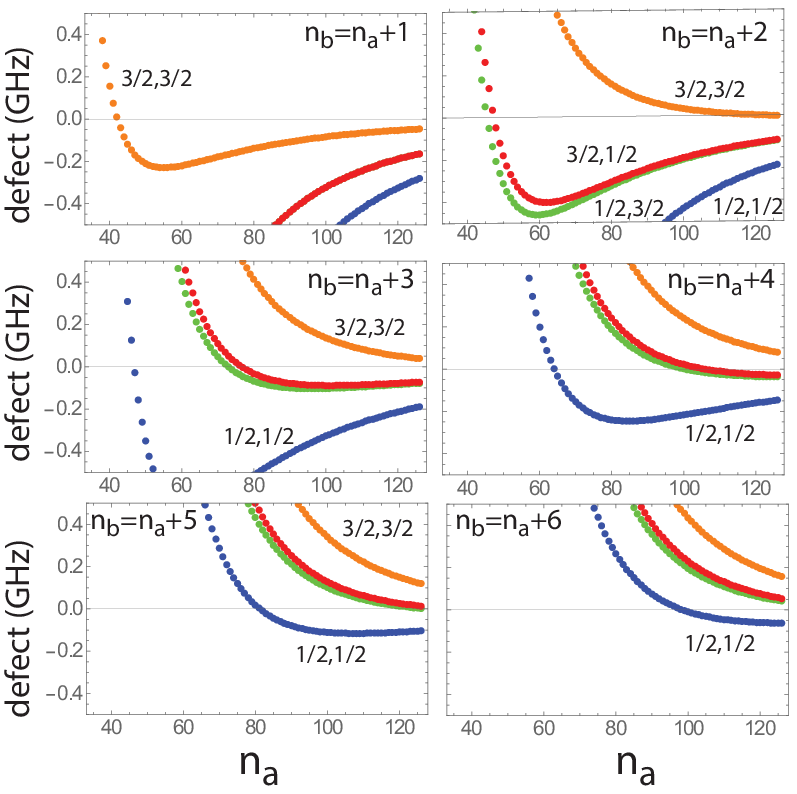}
\caption{(color online) Energy defects for Cs-Cs coupling.  The different curves show the  $(j_{pa},j_{pb})$ channels. }
\label{fig.Cs_Cs_defects}
\end{figure}

Analogous to the Rb-Cs  F\"orster resonances studied in Sec. \ref{sec.RbCs} there are resonances   for Rb-Rb or Cs-Cs interactions.  Figure \ref{fig.Rb_Cs_defects} shows the energy defect for   excitation of
two Rb atoms or two Cs atoms to the same $ns_{1/2}$ state. Even at large $n$ the energy defect is substantial for the dominant channel which limits the interaction strength\cite{Walker2008}. The energy defect can be  reduced using an external field to give a so-called Stark tuned F\"orster resonance, as has been demonstrated experimentally with dc\cite{Ryabtsev2010} or ac\cite{Tretyakov2014} fields.  Alternatively, the interaction strength can be increased substantially, without an electric field,  by exciting each atom to a different $n$ for which there is a F\"orster resonance as shown in Figs. \ref{fig.Rb_Rb_defects}, \ref{fig.Cs_Cs_defects}.
 This type of resonance has been used to advantage in recent atom-photon coupling experiments with Rb atoms\cite{Tiarks2014}. 
Tables \ref{tab.Rbforster}, \ref{tab.Csforster} list the strongest intraspecies resonances. Comparison of the tables for interspecies and intraspecies resonances show that they have similar strength.  

\begin{table*}[!t]
\caption{Dominant F\"orster resonances for Rb-Rb  $ns_{1/2}$ states.   The van der Waals interaction strength and $R_c$ distance are given for the strongest eigenvector for the listed channel. \label{tab.Rbforster}}
    \begin{ruledtabular}
     \begin{tabular}{cc c c c| c c c c}
channel&Rb$\ket{\gamma_a}$ & Rb$\ket{\gamma_\alpha}$& Rb$\ket{\gamma_b}$ & Rb$\ket{\gamma_\beta}$& $\delta/2\pi$ (MHz) & $C_{3,k}~({\rm GHz}~ \mu\rm m^3)$ &  $U_{k,\rm vdW}~({\rm GHz}~ \mu\rm m^6)$ &$R_c~(\mu\rm m)$\\
\hline

4&$38s_{1/2}$ & $38p_{3/2}$ & $39s_{1/2}$ & $38p_{3/2}$ & 4.62 &	-0.621&	-315&	6.39 \\


1&$49s_{1/2}$ & $49p_{1/2}$ & $51s_{1/2}$ & $50p_{1/2}$ & 30.3 &	-1.91&	-214&	4.38 \\

1&$50s_{1/2}$ & $50p_{1/2}$ & $52s_{1/2}$ & $51p_{1/2}$ &-31.5	&-2.08&	244	&4.45\\

2&$65s_{1/2}$ & $65p_{1/2}$ & $67s_{1/2}$ & $66p_{3/2}$ &32.0	&6.30	&-2480	&6.53\\

2&$67s_{1/2}$ & $67p_{1/2}$ & $69s_{1/2}$ & $68p_{3/2}$ &2.84	&7.14	&-36000	&15.3\\

3&$68s_{1/2}$ & $68p_{3/2}$ & $70s_{1/2}$ & $69p_{1/2}$ &4.60&	7.34	&-23500&	13.1\\

3&$69s_{1/2}$ & $69p_{3/2}$ & $71s_{1/2}$ & $70p_{1/2}$ &-6.92	&7.80	&17600	&11.7\\

1&$80s_{1/2}$ & $80p_{1/2}$ & $83s_{1/2}$ & $82p_{1/2}$ & 5.20&	-14.9	&-76400&	15.7\\

1&$81s_{1/2}$ & $81p_{1/2}$ & $84s_{1/2}$ & $83p_{1/2}$ &-2.92&	-15.7	&151000	&19.3\\

1&$82s_{1/2}$ & $82p_{1/2}$ & $85s_{1/2}$ & $84p_{1/2}$ &-10.3&	-16.5	&47400	&12.9\\

     \end{tabular}
     \end{ruledtabular}
     \end{table*}

\begin{table*}[!t]
\caption{Dominant F\"orster resonances for Cs-Cs  $ns_{1/2}$ states.   The van der Waals interaction strength and $R_c$ distance are given for the strongest eigenvector for the listed channel.\label{tab.Csforster}}
    \begin{ruledtabular}
     \begin{tabular}{cc c c c| c c c c}
channel&Cs$\ket{\gamma_a}$ & Cs$\ket{\gamma_\alpha}$& Cs$\ket{\gamma_b}$ & Cs$\ket{\gamma_\beta}$& $\delta/2\pi$ (MHz) & $C_{3,k}~({\rm GHz}~\mu\rm m^3)$ &  $U_{k,\rm vdW}~({\rm GHz}~\mu\rm m^6)$ &$R_c~(\mu\rm m)$\\
\hline

4&$42s_{1/2}$ & $42p_{3/2}$ & $43s_{1/2}$ & $42p_{3/2}$ & 10.2&	-0.867	&-279&	5.49 \\

4&$43s_{1/2}$ & $43p_{3/2}$ & $44s_{1/2}$&$43p_{3/2}$ &-42.4&	-0.961&	82.2	&3.53\\

2&$45s_{1/2}$ & $45p_{1/2}$ & $47s_{1/2}$ & $46p_{3/2}$ &37.7&	1.28 &-86.3	&3.63\\

3&$47s_{1/2}$ & $47p_{3/2}$ & $49s_{1/2}$ & $48p_{1/2}$ &21.7	&1.41	&-184	&4.51\\

1&$64s_{1/2}$ & $64p_{1/2}$ & $68s_{1/2}$ & $67p_{1/2}$ &1.67	&-5.8	&-35700&	16.7 \\

2&$73s_{1/2}$ & $73p_{1/2}$ & $76s_{1/2}$ & $75p_{3/2}$ &-1.88&	10.2	&110000&	19.7\\

3&$76s_{1/2}$ & $76p_{3/2}$ & $79s_{1/2}$ & $78p_{1/2}$ & 7.59&	11.0	&-32100	&12.7\\
3&$77s_{1/2}$ & $77p_{3/2}$ & $80s_{1/2}$ & $79p_{1/2}$ &-3.91&	11.7&	69600	&16.2\\

1&$81s_{1/2}$ & $81p_{1/2}$ & $86s_{1/2}$ & $85p_{1/2}$ &3.86	&-15.7&	-114000	&17.6 \\

     \end{tabular}
     \end{ruledtabular}
     \end{table*}

\section{Angular dependence}
\label{sec.angular}

The description so far has considered only the situation where the atoms are quantized along $\hat z$ which coincides with the molecular axis $\hat R$ connecting atom $a$ to atom $b$. The more general case of $\hat R$ at an angle $\theta$ with respect to $\hat z$ is important for calculating interaction strengths in three dimensional ensembles. As we will see the near spherical symmetry of the interaction which is known for coupling of indistinguishable atomic $ns_{1/2}$ states is substantially modified when we consider distinguishable atomic states.

The angular dependence of the interaction is found by noting that when the molecular axis is rotated relative to a fixed laboratory frame the expansion coefficients $c_{k\ell}$ depend on the rotation angles, as described in \cite{Walker2008}, see also \cite{Glaetzle2014}. 
For the case of initial $ns_{1/2}$ Rydberg states we have 
\begin{eqnarray}
c_{k\ell}(\theta)=\bra{u_{k\ell}(m_a,m_b)}{\bf d}_{m_a,m_{a'}}^{1/2}(\theta){\bf d}_{m_b,m_{b'}}^{1/2}(\theta)\ket{\psi(m_{a'},m_{b'})}\nonumber
\end{eqnarray}
with ${\bf d}_{m_a,m_{a'}}^{1/2}(\theta)$ the reduced Wigner matrix for $j=1/2$ evaluated at angle $\theta$.

The angular behavior of the vanderWaals interaction in channel $k$ is then found by generalizing Eq. (\ref{eq.UpsivdW}) to 
\begin{eqnarray}
U_{\ket{\psi},\rm vdW}(\theta)&=&\sum_\ell |c_{k\ell}(\theta)|^2 U_{k\ell,\rm vdW}\nonumber\\
&=&-f_k(\theta) \frac{C_{3,k}^2}{\hbar\delta_k}\frac{1}{R^6}.
\end{eqnarray}
where $f_k(\theta) = \sum_\ell |c_{k\ell}(\theta)|^2 D_{k\ell}$.

The small $R$ resonant interaction can be treated analytically when a single channel is dominant. The interaction strength in channel $k$  at angle $\theta$ 
is given by Eq. (\ref{eq.Uk}) with $D_{kl}$ replaced by $f_k(\theta)$
\begin{equation}
U_{k}(\theta) = \frac{\hbar\delta_k}{2}\left[1- \left(1+\frac{4f_k(\theta)C_{3,k}^2}{\hbar^2\delta_k^2R^6}\right)^{1/2}\right].
\label{eq.Uktheta}
\end{equation}
In the van der Waals limit we add the interaction from all channels. 
For $s$ states there are two limiting cases of parallel and antiparallel spins. 
For the parallel initial state $\ket{\psi}=\ket{1/2,1/2}$ and $\gamma_a\ne \gamma_b$ we find 
\begin{eqnarray}
c_{k1}(\theta)&=&0,\nonumber\\
c_{k2}(\theta)&=&\sqrt2 \cos(\theta/2)\sin(\theta/2),\nonumber\\
c_{k3}(\theta)&=&\cos^2(\theta/2)\nonumber,\\
c_{k4}(\theta)&=&\sin^2(\theta/2)\nonumber.
\end{eqnarray}
 which results in 
\begin{eqnarray}
f_1&=&\frac{4+6\sin^2(\theta)}{9},\nonumber\\
f_2=f_3&=&\frac{14-6\sin^2(\theta)}{9},\label{eq.fplus}\\
f_4&=&\frac{22+6\sin^2(\theta)}{9}.\nonumber
\end{eqnarray}
For the antiparallel  state $\ket{\psi}=\ket{1/2,-1/2}$ we find 
\begin{eqnarray}
c_{k1}(\theta)&=&1/\sqrt2,\nonumber\\
c_{k2}(\theta)&=&\cos(\theta)/\sqrt2,\nonumber\\
c_{k3}(\theta)&=&-\sin(\theta)/2,\nonumber\\
c_{k4}(\theta)&=&\sin(\theta)/2\nonumber.
\end{eqnarray}
 which results in 
\begin{eqnarray}
f_1&=&\frac{8-6\sin^2(\theta)}{9},\nonumber\\
f_2=f_3&=&\frac{10+6\sin^2(\theta)}{9},\label{eq.fminus}\\
f_4&=&\frac{26-6\sin^2(\theta)}{9}.\nonumber
\end{eqnarray}
For other initial Zeeman states the angular functions $f_k$ will be different than those given here. Identical initial states with $\gamma_a=\gamma_b$ have $f_k$ twice as large as those in Eqs. (\ref{eq.fplus},\ref{eq.fminus}). Although all channels have substantial anisotropy, the total interaction summed over channels, $\sum_k f_k(\theta)$, has no $\theta$ dependence which implies that  the interaction becomes isotropic in the limit of vanishing fine structure splitting between $np_{1/2}-np_{3/2}$.

\begin{figure}[!t]
\centering
\includegraphics[width=8.4cm]{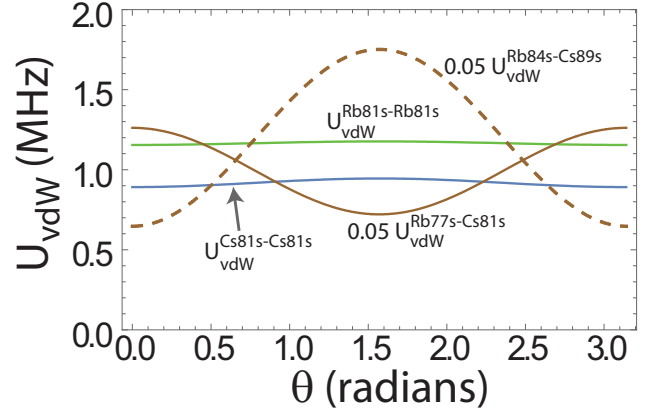}
\caption{(color online) Angular variation of the van der Waals interaction for Rb-Rb, Cs-Cs, and Rb-Cs coupling at $R=12.7~\mu\rm m $. In each case the initial states have $m_a=m_b=1/2$.  }
\label{fig.UvdW}
\end{figure}

We proceed to illustrate these results with some examples. 
Consider  two Cs atoms in the Rydberg state 
$81s_{1/2}\ket{m_a=1/2},81s_{1/2}\ket{m_b=1/2}$ as a function of $\theta$. The interaction is dominated by the coupling $\ket{81s_{1/2}}\ket{81s_{1/2}}\leftrightarrow\ket{81p}\ket{80p}$ with all four fine structure channels contributing. Using $C_{3,k}^2/\delta_k=(-140.4, -237.7, -196.8, -463.0)~ {\rm GHz}~\mu{\rm m}^6$ for $k=1,4$ we find 
\begin{eqnarray}
U_{\rm vdW}^{\rm Cs-Cs}&=&\frac{3740.+225.\sin^2(\theta)}{R^6} ~\rm(GHz)\nonumber
\end{eqnarray}
with $R$ in $\mu\rm m$.
For a pair of Rb atoms in the same $81s_{1/2}$ state we find 
$C_{3,k}^2/\delta_k=(-266.8, -362.1, -334.4, -499.1)~ {\rm GHz}~\mu{\rm m}^6$
and
\begin{eqnarray}
U_{\rm vdW}^{\rm Rb-Rb}&=&\frac{4840.+92.5\sin^2(\theta)}{R^6} ~\rm(GHz)\nonumber.
\end{eqnarray}
The Cs-Cs and Rb-Rb interaction shifts are shown in Fig. \ref{fig.UvdW} as a function of the molecular axis angle $\theta$. We see that both species have comparable interaction strengths that are weakly  anisotropic with an angular variation of about 2\% for Rb and 6\% for Cs. 

The situation can be  markedly different for Rb-Cs. Taking $n_{\rm Cs}=81$
there is a F\"orster resonance with Rb at $n_{\rm Rb}=77$. The interaction is strongly dominated by a single channel $k=3$.  Using $C_{3,k}^2/\delta_k=(-729.0, 638.4, -69020., 345.2)~ {\rm GHz}~\mu{\rm m}^6$ for $k=1,4$ we find 
\begin{eqnarray}
U_{\rm vdW}^{\rm Rb-Cs}&=&\frac{105900.-45330.\sin^2(\theta)}{R^6} ~\rm(GHz).\nonumber
\end{eqnarray}
The interspecies interaction is stronger by about a factor of 20, than for Rb-Rb or Cs-Cs, and is strongly anisotropic with a minimum at $\theta=\pi/2.$ This is because of the dominance of the $k=3$ channel. A different situation arises for the Rb $84s_{1/2}$ - Cs $89s_{1/2}$ resonance. In this case
$C_{3,k}^2/\delta_k=(-136600.,1028.,2025.,683.9)~ {\rm GHz}~\mu{\rm m}^6$ for $k=1,4$. The $k=1$ channel is now dominant and 
\begin{eqnarray}
U_{\rm vdW}^{\rm Rb-Cs}&=&\frac{54300.+92600.\sin^2(\theta)}{R^6} ~\rm(GHz).\nonumber
\end{eqnarray}
As seen in Fig. \ref{fig.UvdW} the interaction now has a maximum at $\theta=\pi/2$ and is strongly anisotropic.  The Rb-Rb or Cs-Cs F\"orster resonances given in Tables \ref{tab.Rbforster},\ref{tab.Csforster} can also be anisotropic depending on which channels dominate.

\section{Quantum nondemolition state measurements with low crosstalk using interspecies coupling}
\label{sec.measurement}

One of the outstanding challenges of neutral atom approaches to quantum computing is the requirement of qubit state measurements without loss, and with low crosstalk to proximal qubits. Such a capability is essential for implementation of quantum error correction. The most widely used approach to  qubit measurements with neutral alkali atoms relies on imaging of fluorescence photons scattered from a cycling transition between one of the qubit states and the strong D2 resonance line\cite{Saffman2010}. Due to a nonzero rate for spontaneous Raman transitions from the upper hyperfine manifold there is a limit to how many photons can be scattered, and imaged, without changing the quantum state. This problem is typically solved by preceding a measurement with resonant ``blow away" light that removes atoms in one of the hyperfine states. The presence or absence of an atom is then measured with repumping light turned on, and a positive measurement result is used to infer that the atom was in the state that was not blown away. 

This method can indeed provide high fidelity state measurements but has several drawbacks. An atom is lost half the time on average, and must be reloaded and reinitialized for a computation to proceed. Atom reloading involves mechanical transport, and thus tends to be slow compared to gate and measurement operations. In addition, error correction would require that a single site in a qubit array can be reloaded, without disturbing proximal qubits. While progress has been made towards this goal\cite{Fortier2007,Khudaverdyan2008,Dinardo2015}, much work remains to be done.

Lossless quantum nondemolition (QND) measurements that leave the atom in one of the qubit states, or at least in a known Zeeman sublevel of the desired hyperfine state, can be performed provided that the measurement is completed while scattering so few photons that the probability of a Raman transition is negligible. This was first done for atoms strongly coupled to a cavity\cite{Boozer2006,Bochmann2010,Gehr2010}, and was subsequently   extended to atoms in free space\cite{Gibbons2011,Fuhrmanek2011,Jau2015}.

Despite these advances, achieving useful QND state measurements in an array of neutral atom qubits remains an outstanding challenge due to the absorption of scattered photons by proximal atoms. Since the resonant cross section for photon absorption is $\sigma=\frac{3}{2\pi}\lambda^2$ and qubits in recent lattice experiments are spaced by $d\sim 5\lambda$\cite{Xia2015,YWang2015} the probability of a scattered photon being absorbed is $\eta_{\rm abs}\sim \sigma/(4\pi d^2)\sim 0.0015.$ If the qubit measurement is performed with a moderately high numerical aperture collection lens of $NA=0.5$ and the optical and detector efficiencies are 50\% the probability of photon detection is $\eta_{\rm det}\sim 0.034$ so that $\eta_{\rm abs}/\eta_{\rm det}\sim  0.04$. This ratio implies that a state measurement based on detection of only a single photon would incur a $\sim 4\%$ probability of unwanted photon absorption at a neighboring qubit. This 4\% error rate is too large to be efficiently handled by protocols for quantum error correction. 

\begin{figure*}[!t]
\centering
\includegraphics[width=16.8cm]{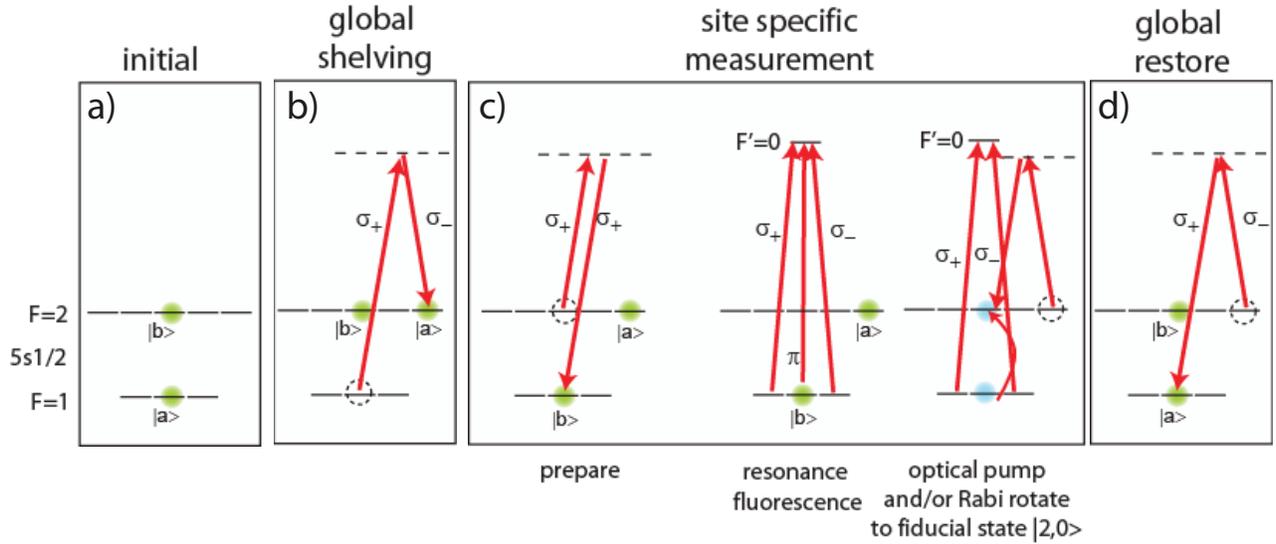}
\caption{\label{fig.global}(color online) Protection protocol for site specific measurements without crosstalk.  Panel a) shows qubits in a superposition of  hyperfine clock states of $^{87}$Rb.  In b) a global shelving operation is performed to map $\ket{a}$ onto $\ket{2,2}$ for all qubits. In c) a site specific mapping of $\ket{b}$ to $\ket{1,0}$ is performed. Resonance fluorescence is then generated using light of all polarizations coupling $5s_{1/2}\ket{f=1}$ to $5p_{3/2}\ket{f'=0}$.  The atom can only decay back to $f=1$ and scattered photons are off-resonant with the shelved atoms in $\ket{2,0}$.   If scattered photons are detected the state is projected into $f=1$ and the measurement result was qubit state $\ket{b}$.  The atom is then pumped into $\ket{1,0}$ using $\sigma_+$ and $\sigma_-$ polarized light, after which it is rotated to $\ket{2,0}$ using a $\pi$ polarized microwave, or Raman light. If no scattered photons are detected the measurement result is qubit state $\ket{a}$ and the atom shelved in $\ket{2,2}$ is rotated back to $\ket{2,0}$ using $\sigma_- - \sigma_+$ Raman light. Finally, in d) the globally protected states are restored back to $\ket{a}$ using Raman light.  }
\end{figure*}

One  approach to solving the crosstalk problem is to protect nearby qubits in states that are dark to scattered resonant photons. This has been used effectively in experiments with trapped ions\cite{Schindler2013}. Such methods are in principle possible with neutral atoms, and an example using a single species  is shown in Fig. \ref{fig.global} for $^{87}$Rb.  Similar ideas  could also be implemented with other species. While the protection protocol can in principle solve the crosstalk problem it has the drawback of requiring both local and global operations, and is thus both complicated to implement and likely to be relatively slow.  Nevertheless this protocol points to an alternative approach using interspecies coupling. The Fig. \ref{fig.global} protocol suppresses crosstalk by placing all but the atom of interest in a dark state with respect to the probe light. Another way of suppressing crosstalk is to use one species for computational qubits and a second species for measurement qubits. Selective mapping of computational to measurement qubits allows us to probe the measurement qubits while keeping the computational qubits in a dark state with respect to the probe light, which is only resonant with the second species. 

\begin{figure}[!t]
\centering
\includegraphics[width=7.4cm]{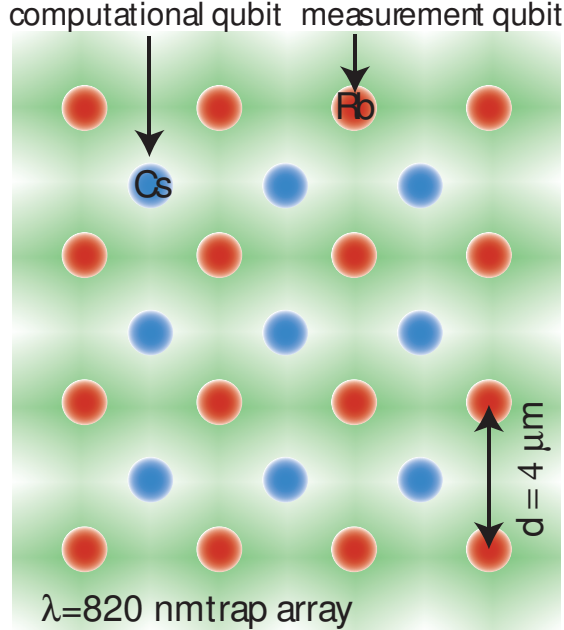}
\caption{\label{fig.array}(color online) Trap array with 820 nm light creates a checkerboard pattern of Cs computational qubits in blue detuned traps and Rb measurement qubits in red detuned traps.     }
\end{figure}

This idea is made explicit using a two-species array as shown in Fig. \ref{fig.array}.   Our approach  is analogous to the demonstrations of quantum logic spectroscopy\cite{Schmidt2005} and entanglement\cite{Tan2015} with two ion species, and builds on earlier ideas of mapping single atoms to ensembles for fast readout\cite{Saffman2005b} as well as the availability of asymmetric Rydberg interactions for creating multiparticle entanglement\cite{Saffman2009b}.   The interspecies protocol   requires fewer operations than in Fig. \ref{fig.global}, and  increases the useful photon rate per atom by a factor of four or more while eliminating crosstalk to other qubits.  

Consider the qubit array shown in Fig. \ref{fig.array}. We assume this is a 2D array of 3D traps for Cs atoms as described in \cite{Piotrowicz2013}, and used for recent experiments with single qubit\cite{Xia2015} and two-qubit\cite{Maller2015a} quantum gates.  We will modify the array slightly by changing the wavelength of the trap light from 780 nm to 820 nm. This is still blue-detuned for Cs atoms which will be trapped at local minima of the optical intensity. The 820 nm light is red detuned for Rb atoms which will be trapped at local maxima of the intensity, forming a checkerboard pattern of alternating Cs and Rb atoms. The lattice period separating atoms of the same species will be   $d= 4~\mu\rm m$, and each Cs atom is surrounded by four Rb atoms at a distance of $d=4/\sqrt{2}=2.8~\mu\rm m$. The large wavelength separation between the Rb resonance lines at 780, 794 nm, the trap light at 820 nm, and the Cs resonance lines at 852, 894 nm allows for independent loading, cooling, control, and measurement of the two species. 

\begin{figure}[!t]
\centering
\includegraphics[width=8.4cm]{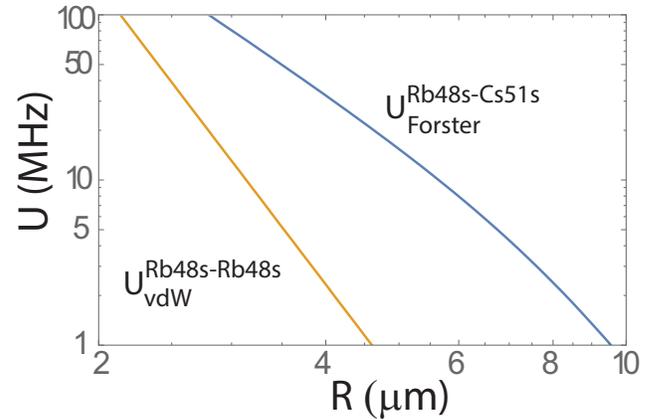}
\caption{\label{fig.Ucross}(color online) Log-log plot of the interspecies coupling strength (upper curve) and Rb-Rb coupling strength (lower curve) for  Rb48s and Cs51s states at $\theta=90~\rm deg.$  The Rb-Cs curve is the full F\"orster interaction of Eq. (\ref{eq.Uk}) for the single dominant channel 3 from Table \ref{tab.forster}. The Rb-Rb curve is the van der Waals interaction summed over all four  channels for $48s48s\leftrightarrow 47p48p$.}
\end{figure}

Let us now choose an interspecies F\"orster resonance that gives strong Rb-Cs coupling and relatively weak Rb-Rb coupling. An example is shown in Fig. \ref{fig.Ucross} for the Rb$48s$-Cs$51s$ channel $3$ resonance from Table \ref{tab.forster}.  Each Cs atom interacts with its nearest neighbor Rb atoms with a  strength of 
$U_{\rm RbCs}/2\pi = 96.8~\rm MHz$ at $R=2.8~\mu\rm m$ and $\theta=\pi/2$. We have assumed that the Cs and Rb atoms are excited to opposite 
$m$ values so we use Eqs. (\ref{eq.Uktheta},\ref{eq.fminus}) to calculate the interaction strength.  In contrast   two Rb atoms interact with a much smaller  $U_{\rm RbRb}/2\pi=2.3~\rm MHz$ at $R=4~\mu\rm m$. To measure the state of a Cs atom qubit we prepare all Rb atoms in the state $\ket{1}_{\rm Rb}=\ket{2,2}$ by optical pumping and then perform the sequence 
\begin{eqnarray} 
{\rm Cs} &:& c_0\ket{0}+c_1\ket{1}\stackrel{\longrightarrow}{\pi}c_0\ket{0}+ic_1\ket{51s}\nonumber\\
{\rm Rb} &:& \ket{\bar 1}\stackrel{\longrightarrow}{\pi}i\ket{\overline{48s}}\stackrel{\longrightarrow}{\pi}-\ket{\bar 0}\nonumber\\
{\rm Cs} &:& c_0\ket{0}+ic_1\ket{51s}\stackrel{\longrightarrow}{-\pi}c_0\ket{0}+c_1\ket{1}\nonumber.
\end{eqnarray}
Provided the Rabi frequency of the Rb Rydberg excitation $\Omega_{\rm Rb}$ is small compared to $U_{\rm RbCs}$ we create the entangled state 
$$
c_0\ket{0}_{\rm Cs}\ket{\bar 0}_{\rm Rb}+c_1\ket{1}_{\rm Cs}\ket{\bar 1}_{\rm Rb}.
$$
The overbar in the Rb kets denotes that this is a multiparticle state of four Rb atoms. 
We then measure the hyperfine state of the Rb atoms. A detector click projects the Cs qubit into $\ket{1}$ and no click projects into $\ket{0}$.

This approach has several advantageous features.  Since each Cs atom is strongly coupled to four nearest neighbors the photon rate can be 
four times greater than for measurement of a single atom.  This reduces the difficulty of obtaining a hyperfine state measurement without suffering a Raman transition.  Furthermore since the state of the Cs qubit is measured using fluorescence light at 780 nm which is far detuned from the Cs resonance lines, crosstalk to other Cs qubits will be negligible. After a measurement the Rb atoms can be rapidly repumped to the $\ket{\bar 1}_{\rm Rb}$ state in preparation for the next measurement. In addition to measurement of a single qubit the Rb atoms could also naturally be used as ancilla qubits for syndrome extraction in quantum error correcting codes. 

We proceed to estimate the measurement fidelity with realistic experimental parameters. When the Cs atom is in state $\ket{0}$ the transfer of the Rb atom between states will be affected by residual couplings $U_{\rm RbRb}$ to nearby atoms. This gives a transfer error for four atoms of \cite{Saffman2009b}
$E=0.72 U_{\rm RbRb}^2/|\Omega_{\rm Rb}|^2.$
The other dominant error is the imperfect blockade when the Cs qubit is in state $\ket{1}$. The   first $\pi$ pulse on the Rb atom creates the state 
$c_{\rm Rb} \ket{48s}$ with 
$$|c_{\rm Rb}|^2=\frac{|\Omega_{\rm Rb}|^2}{|\Omega_{\rm Rb}|^2+U_{\rm RbCs}^2}
\sin^2\left(\sqrt{1+U_{\rm RbCs}^2/|\Omega_{\rm Rb}|^2}\frac{\pi}{2}\right).
$$ 
If we set the Rb Rabi frequency such that $U_{\rm RbCs}/|\Omega_{\rm Rb}|=\sqrt3$ then $c_{\rm Rb}=0$, and there is no state transfer, as desired. This result will be modified slightly by the presence of more than one Rb atom, but an equivalent nulling condition will still exist. For the interaction strengths given above this condition is $\Omega_{\rm Rb}/2\pi=55.9~\rm MHz$ and the transfer error is $E=0.72\times (2.3/55.9)^2 = 0.0012$. 
There is also a spontaneous emission error from the finite lifetime Rydberg states. The Cs qubit is on average Rydberg excited for 
$t=(1/2)(2\pi/\Omega_{\rm Rb} + \pi/\Omega_{\rm Cs})$.  The Rb atom is on average Rydberg excited for $t=(1/2)\pi/\Omega_{\rm Rb}$. The room temperature lifetimes are\cite{Beterov2009,*Beterov2009b} $\tau_{{\rm Rb}48s}=58~\mu\rm s$ and $\tau_{{\rm Cs}51s}=63~\mu\rm s$. Taking $\Omega_{\rm Cs}=\Omega_{\rm Rb}$ we find $P_{\rm se, Rb} = 7.7\times 10^{-5}$ and $P_{\rm se, Cs} = 2.1\times 10^{-4}$.

The largest error is the Rb state transfer at $0.0012$. This small error occurs on average half the time when the Cs atom is in the $\ket{0}$ state and could be reduced 
even further by using a 25\% larger lattice spacing which would increase the $U_{\rm RbCs}/U_{\rm RbRb}$ ratio by a factor of two. It is also likely that   adiabatic or composite pulse sequences can be designed to  minimize the sensitivity to small variations in coupling strength\cite{Beterov2013a}.

Finally we note that the use of two different species, combined with optical tweezers at a wavelength that only perturbs one species at a time, provides a means to move quantum information about in a larger array. This idea was developed for the case of Cs and Li atoms in Ref. \cite{Soderberg2009}. 
In the cited work the entanglement of Cs and Li atoms  was envisioned to occur via short range molecular  interactions. The interspecies Rydberg interaction described here  can in principle be extended to
Cs-Li, or other combinations, with the advantage that interactions can be performed at long range.

\section{Summary}
\label{sec.summary}

We have calculated the interspecies F\"orster interaction between Rb and Cs atoms, as well as F\"orster interactions for Rb-Rb and Cs-Cs where the participating atoms are excited to $ns$ states with different principal quantum numbers. These interactions can be remarkably strong leading to van der Waals interaction strengths of several MHz at  $R=20~\mu\rm m$ for $n<90$.  The strong interactions are of interest for long range coupling between atoms of the same species which has already been demonstrated in Rb ensembles\cite{Tiarks2014}. 

We also propose to use the Rb-Cs interaction for lossless and crosstalk free QND measurements. Needless to say the fidelity of this approach to measurements relies on having high fidelity Rydberg gates available. The current state of the art using the Rydberg blockade interaction, without post selection,  uses a CNOT gate to create  Bell states with a fidelity of 0.73\cite{Maller2015a}. This is much lower than the intrinsic fidelity of the Rb-Cs mapping protocol which we estimate in Sec. \ref{sec.measurement}
to be $\sim 0.001$ with realistic experimental parameters. The two-qubit gate fidelity is therefore the largest roadblock for the protocol analyzed here. On the other hand, there is little interest in QND measurements of single atoms in a qubit array if high fidelity gates are not also available.  When a high fidelity Rydberg gate is demonstrated, the cross entanglement protocol described here may prove valuable for scaling up quantum information tasks with low cross talk. 

\acknowledgments

MS was supported by the IARPA MQCO program through  ARO contract W911NF-10-1-0347, the ARL-CDQI through cooperative agreement W911NF-15-2-0061, the AFOSR MURI, and NSF  award 1521374. IIB acknowledges RFBR grant no. 14-02-00680.


%



\appendix

\section{Channel eigenvalues}

To find the angular factors $D_k(m_a,m_b)$ for channel $k$ and initial Zeeman states $m_a, m_b$ we form the matrix of coefficients 

 \begin{widetext}
\begin{equation}
\bf M
=
\begin{pmatrix}
0&\cdots&0&M_{m_{a1}, m_{b1}}^{-j_{\alpha },-j_{\beta }}&\cdots&M_{m_{a1}, m_{b1}}^{j_{\alpha },j_{\beta }}\\
\vdots&\cdots&\vdots&\vdots&\cdots&\vdots\\
0&\cdots&0&M_{m_{aN_{ab}}, m_{bN_{ab}}}^{-j_{\alpha },-j_{\beta }}&\cdots&M_{m_{aN_{ab}}, m_{bN_{ab}}}^{j_{\alpha },j_{\beta }}\\
%
%
M_{m_{a1}, m_{b1}}^{-j_{\alpha },-j_{\beta }}&\cdots&M_{m_{aN_{ab}}, m_{bN_{ab}}}^{-j_{\alpha },-j_{\beta }}&\hbar \delta_k&\cdots&0\\
\vdots&\cdots&\vdots&\vdots&\ddots&\vdots\\
M_{m_{a1}, m_{b1}}^{j_{\alpha },j_{\beta }}&\cdots&M_{m_{aN_{ab}}, m_{bN_{ab}}}^{j_{\alpha },j_{\beta }}&0&\cdots&\hbar \delta_k
\end{pmatrix}.
\label{eq.M}
\end{equation}

\end{widetext}
The matrix has dimensions $N\times N$ with $N=N_{ab}+N_{\alpha\beta}$ and accounts for the coupling between states with the same value of $m=m_a+m_b$. The laser excited states are referred to as ``initial" states and the dipole coupled Rydberg states as ``target" states. The number of initial states is $N_{ab}=1 +(j_a+j_b)-|m|.$ The number of target states $N_{\alpha\beta}$ is at most  $(2j_{\alpha }+1)(2j_{\beta }+1)$, but may be less than that    due to the requirement that $m_\alpha+m_\beta=m_a+m_b$.  

The nonzero off-diagonal entries are the dipole-dipole matrix elements  
\begin{equation}
M_{m_a,m_b}^{m_{\alpha },m_{\beta }}=-\frac{\sqrt6\, C_{3,k} }{R^3}
\sum_{q=-1}^1 C_{1q1-q}^{20}
C_{j_am_a1q}^{j_{\alpha }m_{\alpha }}C_{j_bm_b1-q}^{j_{\beta }m_{\beta }}
\end{equation}
with $C_{3,k}$ defined in Eq. (\ref{eq.C3k}).
The $N_{\alpha\beta}$ diagonals have  value $\hbar \delta_k$. The eigenvalues and eigenvectors of $\bf M$
give the molecular energies of  Rydberg excited atom pairs  via a single interaction channel as a function of the atomic separation $R$ with the quantization axis along $\hat R$ which points from atom $a$ to atom $b$. 

When the $j$ are half integers, which is the case for alkali atoms, and $N_{ab}=1$ the eigenvalues of $\bf M$ are of the following form. There are $N-2$ degenerate eigenvalues $U=\hbar\delta_k$ which have no $R$ dependence and correspond to admixtures of $\alpha $ and $\beta $ states. The remaining two eigenvalues are 
\begin{widetext}
\begin{equation}
U_{k \pm} = \frac{\hbar\delta_k}{2}\left[1 \pm \left(1+4\frac{\sum_{m_{\alpha }=-j_{\alpha }}^{j_{\alpha }}\sum_{m_{\beta }=-j_{\beta }}^{j_{\beta }}\left(M_{m_a, m_b}^{m_\alpha,m_\beta} \right)^2 } {(\hbar \delta_k)^2} \right)^{1/2}\right].
\end{equation}
At large $R$ the $U_{k-}$ eigenvalue asymptotes to zero and therefore corresponds to $U_k$ of Eq. (\ref{eq.Uk}) whereby we see that
\begin{equation}
D_k(m_a,m_b) = 6\sum_{m_{\alpha }=-j_{\alpha }}^{j_{\alpha }}\sum_{m_{\beta }=-j_{\beta }}^{j_{\beta }}\left(\sum_{q} C_{1q1-q}^{20}
C_{j_am_a1q}^{j_{\alpha },m_{\alpha }}
C_{j_bm_b1-q}^{j_{\beta },m_{\beta }} \right)^2.
\end{equation}
\end{widetext}
When $\gamma_a=\gamma_b$ the eigenvalue is $2D_k$. 
When $N_{ab}>1$ the eigenvectors are superpositions of $\ket{m_a,m_b}$ states and it is not possible to give  compact expressions for $U_k, D_k$. In these cases we extract the $D_k$ from the calculated eigenvalues by comparison with Eq. (\ref{eq.Uk}).

\end{document}